\pdfoutput=1 
\documentclass[prl,twocolumn,preprintnumbers,superscriptaddress,showpacs]{revtex4-1}
\usepackage{CJKutf8}
\usepackage{graphicx}
\usepackage{bm}
\usepackage{amsmath}
\usepackage{amssymb}
\usepackage[colorlinks,pdfstartview=FitH]{hyperref}
\usepackage{comment}
\hypersetup{linkcolor=blue,citecolor=blue,filecolor=black,urlcolor=blue}
\newcommand\sect[1]{\emph{#1.}---}

\newcommand\pd{\partial}

\newcommand\Hfunction{{\cal H}}
\newcommand\Hcurrent{{\cal H}}
\newcommand\HCurrent{{H}}

\begin{document}
\begin{CJK}{UTF8}{gbsn}

\preprint{EFI-15-7}
\author{Jing-Yuan~Chen (陈静远)}
\affiliation{Kadanoff Center for Theoretical Physics, University of 
Chicago, Chicago, Illinois 60637, USA}
\author{Dam T.~Son}
\affiliation{Kadanoff Center for Theoretical Physics, University of 
Chicago, Chicago, Illinois 60637, USA}
\author{Mikhail A.~Stephanov} 
\affiliation{Physics Department, University of Illinois at Chicago, Chicago, 
Illinois 60607, USA}

\title{Collisions in Chiral Kinetic Theory}
\date{February 2015}

\begin{abstract}
  Using a covariant formalism, we construct a chiral kinetic theory
  Lorentz invariant to order $\mathcal
    O(\hbar)$ which includes collisions. 
  We find a new contribution to the particle number current due to
 the side jumps required by
  the conservation of angular momentum during collisions.
  We also find a conserved symmetric stress-energy tensor as well as
  the $H$-function obeying Boltzmann's $H$-theorem.  We demonstrate their use
  by finding a general equilibrium solution and the values of the anomalous
  transport coefficients characterizing chiral vortical effect.
\end{abstract}
\maketitle
\end{CJK}

\sect{Introduction}%
The role of chiral anomalies in the collective dynamics has attracted
considerable attention recently.  It has been known for some
time~\cite{PhysRevD.20.1807,PhysRevD.22.3080} that a chiral
medium in a magnetic field or in rotation can respond by a current
along the field or the rotation axis -- the chiral magnetic or chiral
vortical effects (CME and CVE).  More recently, the effects of anomalies
in medium have come up in several different experimental and
theoretical contexts.
Charge-dependent correlations which may be driven by the
CME~\cite{Fukushima:2008xe}  have been observed in heavy-ion
collisions.
The CVE in hydrodynamics has been discovered using gauge/gravity duality
\cite{Erdmenger:2008rm,Banerjee:2008th}. Later, both CME and
CVE were shown to be universally required by the second law of
thermodynamics~\cite{Son:2009tf}.  The recent discovery of ``3D
graphene'' \cite{2013arXiv1309.7892N,2014Sci...343..864L} and the
possible observation~\cite{Li:2014bha} of the CME-induced negative
magnetoresistance~\cite{Son:2012bg} have opened a new experimental
frontier for investigating physical consequences of anomalies.


Despite the recent progress, the role of anomaly in kinetic theory
has not been completely understood.  Kinetic theory is essential for
the understanding of nonequilibrium dynamics and is applicable when
external fields are weak and collisions are rare, so
that each particle moves along its classical trajectory most of the
time.  Recent literature focuses on the kinetic theory
without collisions.  It was shown~\cite{Son:2012wh}
that anomaly is encoded in the momentum-space Berry curvature, and the
action for such a motion has been derived
microscopically~\cite{Stephanov:2012ki,Son:2012zy}.
Although the action and the equations of motion are not manifestly
relativistic, a hidden Lorentz
invariance, involving nontrivial modifications of Lorentz
transformations, has been found up to order ${\cal
O}(\hbar)$~\cite{Chen:2014cla}.  
Such modifications lead to side
jumps necessary to ensure angular momentum conservation in collisions.
However, the corresponding modifications to the collision term have
not been found so far.  

In this Letter we supply this so far missing important piece of the
theory.  First, we introduce a simple covariant formalism allowing us
to demonstrate Lorentz invariance in an elegant and straightforward
manner.  We then discover that the side jumps not only make
the collision integral
nonlocal, but also require nontrivial contributions to the particle number,
 energy-momentum and entropy currents.
We prove the validity of Boltzmann's
$H$-theorem, guaranteeing relaxation to equilibrium.  
We determine the values of the CVE transport
coefficients from the kinetic theory.  With the goal of understanding
the CVE, we focus on the physics of collisions without
external electromagnetic fields, which will be considered elsewhere.

\sect{Spin and relativity of particle worldline}%
First of all we need
 to generalize the side jump found in Ref.~\cite{Chen:2014cla} to
{\em finite} Lorentz transformations.  Let us consider the
angular momentum tensor of a relativistic
spinning particle,
\begin{equation}
  \label{eq:J}
  J^{\mu\nu}=x^\mu p^\nu - x^\nu p^\mu + S^{\mu\nu},
\end{equation}
where $S^{\mu\nu}$ is the spin.  In
relativistic classical mechanics the separation between orbital
motion and internal rotation as well as the definition of
the center of mass are ambiguous. One can shift $x^\mu$ by
$\Delta^\mu$ and, simultaneously, $S^{\mu\nu}$ by $\Delta^\nu p^\mu
- \Delta^\mu p^\nu$ without changing $J^{\mu\nu}$. 
To define unambiguously the particle
position $x^\mu$, one needs to
impose a gauge-fixing condition on $S^{\mu\nu}$.  For a {\em
  massless} particle ($p\cdot p=0$), the only Lorentz-covariant
condition $p_\mu S^{\mu\nu}=0$ is not sufficient -- leaving
residual shifts $\Delta$ satisfying
$\Delta\cdot p=0$.  To fix the gauge completely one chooses an arbitrary frame and
uses its 4-velocity  $n$ to impose
\begin{equation}
  \label{eq:nS}
  n_\nu S^{\mu\nu} = 0,
\end{equation}
i.e., one requires that $S^{\mu\nu}$ has only spatial components in
the frame $n$.
Together the two conditions $p_\mu S^{\mu\nu}=n_\nu S^{\mu\nu}=0$
fix $S^{\mu\nu}$ in terms of $n$ and $p$ up to an overall factor
\begin{equation}
  \label{eq:Sn}
  S_n^{\mu\nu} = \lambda\, \frac{\epsilon^{\mu\nu\alpha\beta} p_\alpha
    n_\beta}{p\cdot n}\,.
\end{equation}
In the frame where $n^\mu=(1,\bm 0)$,
$S_n^{ij}=\lambda\epsilon^{ijk} p^k/|\bm p|$ and thus $\lambda$ is the helicity
of the particle.

The frame dependence of the
spin tensor $S_n^{\mu\nu}$ in
Eq.~(\ref{eq:Sn}) implies that the particle position $x$
also depends on the frame choice, so that
the total angular momentum in Eq.~(\ref{eq:J}) does not.
This means that if one changes the frame from $n$ to $n'$ the position shifts,
$x'=x+\Delta_{nn'}$, so that
\begin{equation}
  \label{eq:S'-S}
  S_{n'}^{\mu\nu} - S_n^{\mu\nu} =  p^\mu \Delta_{nn'}^\nu -  p^\nu \Delta_{nn'}^\mu \,.
\end{equation}
Dotting this equation with $n_\nu$ and choosing the point on the shifted world line
where the shift is spatial in frame~$n$, $\Delta_{nn'}\cdot n=0$, we find
\begin{equation}
  \label{eq:Delta-lambda}
  \Delta_{nn'}^\mu = - \frac{S_{n'}^{\mu\nu}n_\nu}{p\cdot n} = 
\lambda\, \frac{\epsilon^{\mu\alpha\beta\gamma} p_\alpha
    n_\beta n'_\gamma}{(p\cdot n)(p\cdot n')}\,.
\end{equation}
This is the finite generalization of the infinitesimal side jump found
in Ref.~\cite{Chen:2014cla}. Finite side jumps have been also recently
considered in Refs.~\cite{Duval:2014cfa,Stone:2015kla}.

\sect{Collisionless current}%
We now consider kinetic theory, where the system is characterized by
the phase space particle density $f$.
As the particle positions depend on the frame, so will $f$.  Let us
first ignore collisions, in which case $f$ is constant along the
world lines. Assuming
$f$ and $f'$ in two frames
$n$ and $n'$ are related by $f'(x')=f(x)$, we find to linear order in
$\hbar$, with $\lambda\sim{\cal O}(\hbar)$,
\begin{equation}
  \label{eq:fdiff}
  f'(x) - f(x) = - \Delta\cdot\partial f 
\qquad\mbox{(collisionless)},
\end{equation}
where $\Delta\equiv\Delta_{nn'}$.

The naive phase-space current $p^\mu f$ is thus not a Lorentz vector, since
$f$ is not a scalar field (its value at a given point depends on the frame).
One part of the solution was found in
Ref.~\cite{Chen:2014cla}: the covariant current must include a
magnetization contribution which, in the classical picture, is caused by the
intrinsic rotation of the particles. In our covariant notations
\begin{equation}\label{eq:jnoc}
  j^\mu = p^\mu f + S^{\mu\nu}\partial_\nu f
\qquad\mbox{(collisionless)},
\end{equation}
where $S^{\mu\nu}=S_n^{\mu\nu}$.
Both $f$ and $S^{\mu\nu}$ transform non-trivially under the frame
change $n\to n'$ according to Eqs.~(\ref{eq:S'-S}) and
(\ref{eq:fdiff}) but, after cancellations,
\begin{equation}
  \label{eq:diff}
j'^\mu - j^\mu =
- \Delta^\mu(p\cdot\partial f) 
\qquad\mbox{(collisionless)}.
\end{equation}
Thus the current in Eq.~(\ref{eq:jnoc}) is
frame independent in {\em collisionless} kinetic theory where
$p\cdot\partial f=0$.

Collisions will make the current in Eq.~(\ref{eq:jnoc})
frame dependent. To solve this problem, we have to step back
and try to understand what contribution to the current we may have missed.
 Equation~(\ref{eq:diff}) hints that it 
is related to the side jump and is proportional to the collision rate.

\sect{Collisions and jump current}%
%
%
%
 Let us look at the collisions more closely and,
for simplicity, consider elastic $2\to2$ collisions. From the
classical point of view, such collisions involve
2 incoming and 2 outgoing world lines.
It is convenient to think of incoming particles 
as being annihilated and outgoing particles 
as being created in that
process.  For particles without spin, we can assume that 
all 4 annihilation/creation
events happen at the same spacetime point $x$. The continuity of the
particle current is obvious in this case.

However, for a spinning particle, this cannot remain true in all frames,
because that would contradict conservation of angular
momentum~\cite{Chen:2014cla}.  We assume here that
for each given collision kinematics there is a special
frame --- the ``no-jump frame'' $\bar n$ --- in which
all four particle worldlines converge to 
one spacetime point as in the spinless case.
The natural choice for this special frame is the center of mass frame:
$\bar n = 
{(p_A+p_B)}/{\sqrt s}$, where $p_A$ and $p_B$ are the momenta of the
incoming particles.
To ensure continuity of the current
in a given (lab) frame $n\neq\bar n$ we must include a ``jump current'' associated
with the spacelike motion of each participant particle
between the common collision spacetime point  $x$ 
and the particle's annihilation/creation point in the lab frame,
$x+\bar\Delta$, where from Eq.~(\ref{eq:Delta-lambda})
\begin{equation}
  \label{eq:Delta-bar}
  \bar\Delta^\mu \equiv \Delta_{\bar n n}^\mu 
= \lambda\, \frac{\epsilon^{\mu\alpha\beta\gamma} p_\alpha
    \bar n_\beta  n_\gamma}{(p\cdot n)(p\cdot\bar n)}\,.
\end{equation}
This tunneling-like motion of the particle during the collision would be from
$x+\bar\Delta$ to $x$ if the particle is incoming, or the reverse
if it is outgoing.  Weighing by the probability of the collision with each
given kinematics, we are led to consider the current
\begin{equation}
  \label{eq:jcoll}
 j^\mu = p^\mu f +
    S^{\mu\nu}\partial_\nu f \ 
+ \int\limits_{BCD}C_{ABCD}\,
\bar\Delta^\mu ,
\end{equation}
where we introduced short-hand notations for the usual Lorentz invariant
integration over the phase space of the particles $B$, $C$, and $D$:
\begin{equation}
  \label{eq:int}
  \int \frac{d^4p_B}{(2\pi)^3}
{ 2}{\delta(p_B\cdot p_B)\theta(n\cdot p_B)} \equiv \int_{p_B} \equiv \int_B ,
\end{equation}
etc.\ and for the collision kernel
\begin{equation}
  \label{eq:CW}
  C_{ABCD}\equiv W_{CD\to AB}-W_{AB\to CD}\,,
\end{equation}
where $W$ is the rate of collisions with given momenta
$p_A\equiv p$, $p_B$, $p_C$ and $p_D$. The signs of the
two terms reflect the directions of the jump depending on whether
$A$ is incoming or outgoing.

Let us now check Lorentz covariance of the current $j^\mu$ in
Eq.~(\ref{eq:jcoll}) by considering a different frame $n'$, as we did
before in Eq.~(\ref{eq:diff}). Comparing 4-vectors $j'^\mu$ and
$j^\mu$ we find this time
\begin{multline}
  \label{eq:diff-col}
  j'^\mu - j^\mu 
= p^\mu (f' - f + \Delta \cdot \partial f) - \Delta^\mu(p\cdot\partial f) \ 
\\+ \int\limits_{BCD}C_{ABCD}\,
(\Delta_{\bar n n'}^\mu-\Delta_{\bar n n}^\mu).
\end{multline}
The last term can be transformed using Eqs.~(\ref{eq:Delta-lambda}) and~(\ref{eq:S'-S})
\begin{equation}
  \label{eq:DeltaDelta}
  \Delta_{\bar n n'}^\mu-\Delta_{\bar n n}^\mu = 
-\frac{\left(S'-S\right)^{\mu\nu}\bar n_\nu}{p\cdot\bar n}
= 
\Delta_{nn'}^\mu - p^\mu\ \frac{\Delta_{nn'}\cdot\bar n}{p\cdot\bar n}\,.
\end{equation}
The meaning of Eq.~(\ref{eq:DeltaDelta}) 
is straightforward:
the jump from $\bar n$ to $n'$
equals the jump from $\bar n$ to $n$ plus the jump from
$n$ to $n'$ ($\Delta$) up to a shift along the world line [the
last term in Eq.~(\ref{eq:DeltaDelta})]. 

Substituting into Eq.~(\ref{eq:diff-col}) we observe that $\Delta^\mu$
is independent of the integration variables $p_B$, etc., and thus can
be taken outside of the integration. The remaining integral
coincides with the collision rate
\begin{equation}
  \label{eq:Cint}
   {\cal C}(x;p\equiv p_A) = \int\limits_{BCD} C_{ABCD}\,,
\end{equation}
and, since by kinetic equation $p\cdot\partial f ={\cal C}[f]+{\cal O}(\hbar)$, this term
cancels the $\Delta^\mu (p\cdot \partial f)$ term in
Eq.~(\ref{eq:diff-col}) to order $\hbar$.

To cancel the last term in Eq.~(\ref{eq:DeltaDelta}) substituted into Eq.~(\ref{eq:diff-col}) the
distribution function must transform under the Lorentz transformation (in
addition to the shift of the argument by $\Delta$ in Eq.~(\ref{eq:fdiff})) as
\begin{equation}
  \label{eq:f}
  f' - f = -\Delta\cdot\partial f\  +
\int\limits_{BCD} C_{ABCD}\,
\frac{\Delta\cdot\bar n}{p\cdot\bar n}.
\end{equation}
The additional term in Eq.~(\ref{eq:f}) compared to Eq.~(\ref{eq:fdiff})
 accounts for the colliding particles
undergoing the side jumps. Thus we verified that
the phase-space current $j^\mu$ in Eq.~(\ref{eq:jcoll}) is Lorentz 
covariant provided $f$ transforms as Eq.~(\ref{eq:f}) and $C_{ABCD}$ is
Lorentz invariant.

\sect{Collision kernel}%
Using Eq.~(\ref{eq:jcoll}) and $n\cdot\bar\Delta=0$ 
we see that
$  \label{eq:fj}
  f = {n\cdot j}/{n\cdot p}\,,
$
i.e., naturally, the time component of the current, $j^0$, divided by the
particle energy in the frame~$n$. Since the collision
probability  $ W_{AB\to CD}$ must be a Lorentz scalar, i.e., independent of
$n$, the frame-dependent $f$ cannot directly determine
 $W_{AB\to CD}$ as in $|M|^2 f_Af_B (1-f_C)(1-f_D)$.
Instead, we must use
the distribution function in a frame, associated with the collision itself.
The most natural choice
is the ``no-jump'' frame $\bar n$
\begin{equation}
\bar f = \frac{\bar n\cdot j}{\bar n\cdot p}\,.\label{eq:fbar}
\end{equation}

Now, with the $n$-independent  distribution
function in Eq.~(\ref{eq:fbar}) we can write~\footnote{Although this is, by far,
  the most natural form for the collision probability, satisfying
  nontrivial conditions,
we must still consider this form of the
  collision integral as an educated guess or a conjecture. The correct
  form of the collision term would be found by deriving the kinetic
  theory from underlying field theory, which is beyond present scope.
  Here we only establish that such a consistent theory can be written down in
  principle.}
\begin{multline}
  \label{eq:W}
  W_{AB\to CD}[\bar f] = \frac{1}{2!}|M(s,t)|^2
  (2\pi)^4\delta^4(p_A+p_B-p_C-p_D)\,
\\\times
\bar f_A\bar f_B (1-\bar f_C)(1-\bar f_D) ,
\end{multline}
where factor ${1}/{2!}$ accounts for the indistinguishability of the
outgoing particles.
Using the Lorentz covariant $j^\mu$ and ${\cal C}$, we can write 
a Lorentz invariant chiral kinetic theory
with collisions
\begin{equation}
  \label{eq:djC}
  \partial\cdot j = {\cal C}[\bar f] ,
\end{equation}
where $j^\mu$ is given by Eq.~(\ref{eq:jcoll}) and $\cal C$ is given by
Eqs.~(\ref{eq:Cint}),~(\ref{eq:CW}),~(\ref{eq:W}) with $\bar f$ from
Eq.~(\ref{eq:fbar}). 

Using Eqs.~(\ref{eq:jcoll}), (\ref{eq:Cint}) and the transformation of
$f$ in Eq.~(\ref{eq:f}) we can also rewrite
Eq.~(\ref{eq:djC}) 
as
\begin{multline}
  \label{eq:pdf2}
    p\cdot\partial f 
=  \int\limits_{BCD} 
C_{ABCD}[f]
\\\times
\biggl(1
-
\int\limits_{B'C'D'}\frac{\partial}{\partial f}
C_{AB'C'D'}
\frac{\bar\Delta\cdot\bar n'}{p\cdot\bar n '}\,
\biggr)
\,,
\end{multline}
where $\bar n'$ is the no-jump frame of the collision
$AB'\leftrightarrow C'D'$.
In the form of Eq.~(\ref{eq:pdf2}) Lorentz invariance
is not manifest as in Eq.~(\ref{eq:djC}), but the collision
kernel is expressed solely in terms of the distribution function $f$ in the
lab frame. Equations.~(\ref{eq:djC}) and~(\ref{eq:pdf2}) are equivalent to
linear order in $\hbar$.

\sect{Conserved currents}%
Since the underlying quantum theory of Weyl fermions
 is invariant under CPT, we must take into account antiparticles, which 
also participate in collisions. These can be easily
 incorporated by considering the particle charge $q=\pm 1$ as an
 additional discrete index of the distribution function $f(x,p,q)$,
 indices $A$, $B$, etc. as composite indices $A=(p_A,q_A)$, etc. and
 accompanying integration over $p$ by summation over $q$. CP
 invariance implies $\lambda=q|\lambda|$. The net current of $q$ is
 given by 
 \begin{equation}
   \label{eq:Jq}
   J_q^\mu = \sum_q \int_p q j^\mu ,
 \end{equation}
and its conservation, $\partial_\mu J_q^\mu=0$, follows from
Eq.~(\ref{eq:djC}) and the charge conservation in a collision: $\sum_q\int_p q {\cal C}=0$.

Similarly, one can show that the following covariant symmetric
(and traceless) tensor
\begin{equation}
  \label{eq:Tmunu}
  T^{\mu\nu}=  \sum_q\int_p \frac12(\,p^\mu j^\nu + p^\nu j^\mu\,)
\end{equation}
is conserved $\pd_\nu T^{\mu\nu}=0$ due to the
energy-momentum and angular momentum conservation in the collisions.

\sect{Entropy current and $H$-theorem}%
%
An important property of kinetic theory is the existence of the entropy---a
functional of $f$ which does not decrease with time. This
is known as the $H$-theorem, which guarantees that the
system relaxes to equilibrium. To prove the $H$-theorem we
need to find the corresponding covariant current $\HCurrent^\mu$ whose
divergence is non-negative.

First let us generalize current $j^\mu$ to a current describing
advection of a  generic, for now, quantity ${\cal H}$ which is a
function of the distribution function $f$. Following the same steps as
for the current $j^\mu$ in Eq.~(\ref{eq:jcoll}) we find that the
following current
\begin{equation}
  \label{eq:Phi-LI}
  \Hcurrent^\mu =   p^\mu\Hfunction 
+ S^{\mu\nu}\partial_\nu \Hfunction\ 
+\int\limits_{BCD}C_{ABCD}\,
\bar\Delta^\mu\,\frac{\partial\Hfunction}{\partial f}
\end{equation}
does not depend on the choice of the frame $n$ to linear order in
$\hbar$.

Using Eq.~(\ref{eq:pdf2}) one can also show that
\begin{equation}
  \label{eq:dPhi-new}
    \partial_\mu \Hcurrent^\mu
=
\int\limits_{BCD}C_{ABCD}[\bar f]\,\frac{\partial\Hfunction}{\partial \bar f}
\,.
\end{equation}

Furthermore, using the $AB\leftrightarrow CD$ symmetry of the amplitude $|M|$ in
Eq.~(\ref{eq:W}) we can write Eq.~(\ref{eq:CW}) as
\begin{equation}
  \label{eq:CW-r}
  C_{ABCD}=W_{AB\to CD}\,(r-1),
\end{equation}
where
\begin{equation}
  \label{eq:W/W}
  \frac{W_{CD\to AB}}{W_{AB\to CD}} 
= \frac{\bar f_C \bar f_D (1-\bar f_A) (1-\bar f_B)}
{\bar f_A \bar f_B (1-\bar f_C) (1-\bar f_D)} \equiv r,
\end{equation}
and express the current  $\HCurrent^\mu\equiv\int_p \Hcurrent^\mu$ 
as
\begin{equation}
  \label{eq:dPhi-new-2}
    \partial_\mu \HCurrent^\mu
=
\int\limits_{ABCD}W_{AB\to CD}\,(r-1)\,
\frac{\partial\Hfunction}{\partial \bar f_A}\,.
\end{equation}

Now, choosing $\Hfunction$ so that 
${\partial\Hfunction}/{\partial f} = \ln[(1-f)/f]$, i.e.,
\begin{equation}
  \label{eq:phi-log}
  \Hfunction = f\ln\frac{1}{f} + (1-f)\ln\frac{1}{1-f}\,,
\end{equation}
and using the $A\leftrightarrow B$ and $C\leftrightarrow D$
symmetry of  $|M|$ in Eq.~(\ref{eq:W})
we can write for the
divergence of the entropy current $\HCurrent^\mu$ in Eq.~(\ref{eq:dPhi-new-2})
\begin{equation}
  \label{eq:dS}
  \partial_\mu {\HCurrent^\mu}
= \frac14 \int\limits_{ABCD}W_{CD}\,(r-1)\,\ln r \ge 0.
\end{equation}
The rate of entropy production $\partial\cdot \HCurrent$ vanishes when $r=1$.

\sect{Equilibruim}%
\label{sec:cve}
Let us denote, for convenience,
\begin{equation}
  \label{eq:g-f}
  g(f) \equiv \ln\frac{1-f}{f},
\end{equation}
i.e., $f(g)=1/(\exp g + 1)$. 
In terms
of $\bar g \equiv g(\bar f)$, 
the ratio $r$ in Eq.~(\ref{eq:W/W}) is given by
\begin{equation}
  \label{eq:r-g}
  r= \exp(\bar g_A + \bar g_B - \bar g_C - \bar
  g_D).
\end{equation}
The collision kernel in Eq.~(\ref{eq:CW-r}) vanishes if $r=1$
(detailed balance),
 which happens if $\bar g$ is a
linear combination of quantities conserved in the collision (energy, momentum,
angular momentum and charge), i.e.,
\begin{equation}
  \label{eq:g=beta-mu-S}
  g(\bar f_{\rm eq}) = p\cdot \bar U + \frac12 \bar S^{\alpha\beta}\bar\Omega_{\alpha\beta}
  - q \bar Y,
\end{equation}
where $\bar S=S_{\bar n}$ is the spin tensor in the no-jump frame
(orbital momentum is zero), $q$ is the charge and $\bar U$, $\bar Y$ and
$\bar\Omega_{\alpha\beta}=-\bar\Omega_{\beta\alpha}$ are coefficients
(possibly $x$-dependent). The distribution function
$f_{\rm eq}$ in another frame $n$ unrelated to the collision kinematics can be obtained by
transformation~(\ref{eq:f}), according to which
 $g=\bar g -
(\bar\Delta\cdot\partial)\bar g$ (since $C_{ABCD}[\bar f_{\rm eq}]=0$).
Also expressing $\bar S$ in Eq.~(\ref{eq:g=beta-mu-S}) in terms of $S$
using Eq.~(\ref{eq:S'-S}) we can then write for $f_{\rm eq}$:
\begin{equation}
  \label{eq:g=beta-mu-S-nobar}
  g(f_{\rm eq}) = p\cdot U + \frac12 S^{\mu\nu}\Omega_{\mu\nu}
  - q Y,
\end{equation}
where $U_\alpha = \bar U_\alpha + (\bar
\Omega_{\alpha\beta}-\partial_\alpha \bar U_\beta))
\bar\Delta^\alpha$, $\Omega=\bar\Omega$ and $Y=\bar Y -
(\bar\Delta\cdot\partial)\bar Y$. The dependence on the collision
kinematics via vector $\bar n$ (in $\bar\Delta$ according to
Eq.~(\ref{eq:Delta-bar})) drops out, as it must, if $\partial_\alpha
\bar U_\beta=\bar \Omega_{\alpha\beta}$ and $\bar Y={\rm const}$,
which also means $U=\bar U$ and $Y=\bar Y$. The distribution in
Eq.~(\ref{eq:g=beta-mu-S-nobar}) describes a rotating (shear-free) fluid. It
is easy to check that $f_{\rm eq}$ given by Eq.~(\ref{eq:g=beta-mu-S-nobar})
solves kinetic equation (\ref{eq:djC}). In the conventional notations
$U =\beta u$, where $\beta=\sqrt{U\cdot U}$ and $Y=\beta\mu$.

\sect{Chiral vortical effect}%
Now, for the rotating distribution in Eq.~(\ref{eq:g=beta-mu-S-nobar}), we
can calculate the number current $J_q^\mu$ in Eq.~(\ref{eq:Jq}) using
Eq.~(\ref{eq:jcoll}). It is convenient to express the
distribution in the local comoving frame, i.e., choose
$n=u$ (we have not relied on $n$ being coordinate independent). 
To linear order in gradients we find
\begin{equation}
  \label{eq:j-omega}
  J_q^\mu 
=
n_q u^\mu + \xi\omega^\mu,
\end{equation}
with
$\omega^\mu\equiv\frac12\epsilon^{\mu\alpha\beta\gamma} u_\alpha\partial_\beta
u_\gamma$,
$  n_q \equiv \sum_q\int_p (p\cdot u) q f_0$  and
\begin{equation}
  \label{eq:xi}
\xi \equiv \beta\, \sum_q\int_p (p\cdot u) q\lambda
\left(\!-\frac{df_0}{ dg}\right)
=
\frac{\mu^2}{4\pi^2} + \frac{ T^2}{12}\,,
\end{equation}
where  $-df_0/dg = f_0(1-f_0)$, $T=1/\beta$ and
 $f_0$ is the Fermi-Dirac distribution to zeroth order in gradients, i.e.,
$g(f_0)=\beta(p\cdot u - q\mu)$. 

Similarly, for the stress energy tensor in Eq.~(\ref{eq:Tmunu}) we
find
\begin{equation}
  \label{eq:Tcve}
  T^{\mu\nu} = 
w u^\mu u^\nu - p g^{\mu\nu} + \xi_T(\omega^\mu u^\nu + \omega^\nu u^\mu),
\end{equation}
where $w$ and $p$ are the usual expressions for the enthalpy and
pressure of the Weyl gas and
\begin{equation}
  \label{eq:xiT}
  \xi_T =  \frac23\beta\, \sum_q\int_p (p\cdot u)^2 \lambda
  \left(\!-\frac{df_0}{ dg}\right)
= \frac{\mu^3}{6\pi^2} + \frac{\mu T^2}{6}\,,
\end{equation}
which is twice the result found in Ref.~\cite{PhysRevD.20.1807} due to
the contribution of the spin coupling to vorticity.
For the entropy current in Eq.~(\ref{eq:Phi-LI}) we find
\begin{equation}
  \label{eq:HCVE}
  \HCurrent^\mu = s u^\mu + \xi_H\omega^\mu,
\end{equation}
where $s=\beta(w-\mu n)$ and
\begin{equation}
  \label{eq:2}
  \xi_H = \frac32 \beta\,\xi_T - \beta\mu\, \xi
= \frac{\mu T}{6}\,.
\end{equation}
One can check that these results agree with the
general form found in Ref.~\cite{Neiman:2010zi} required by
the second law of thermodynamics.


\acknowledgments
This work is
supported, in part, by a Simon Investigator grant from the Simons
Foundation and the US DOE grants Nos.\ DE-FG0201ER41195 and
DE-FG02-13ER41958.

\bibliography{Chiral_Collision}

\end{document}